# A Comparative study of catalytic activity and lifetime of novel micro-meso porous catalysts in MTO process

**Seyed Hesam Mousavi, Shohreh Fatemi\*, Marjan Razavian**

*School of Chemical Engineering, College of Engineering, University of Tehran, P.O. Box 11365 4563, Tehran, Iran*

*E-mail\*: Shfatemi@ut.ac.ir*

**Abstract**

Recently, two kinds of mesoporous catalysts with high propylene selectivity in propane dehydrogenation (PDH) process has been successfully synthesized. The first proposed catalyst is SAPO-34 molecular sieve with hierarchical tuned nanostructure. The second catalyst is a novel bi-phase SAPO-34/ZSM-5 zeolite hierarchical composite utilized with TPA-SAPO-34 exchanged core cystals being wrapped by ZSM-5 zeolite particles. In this contribution, the physico-chemical properties of the catalysts were analyzed by XRD, FESEM and N2 adsorption-desorption techniques and their catalytic activity and life time were investigated in MTO process. The results show that hierarchical SAPO-34 has a significant lifetime and selectivity to the light olefins compared with SAPO-34/ZSM-5. It shows a full conversion of MeOH during the first 200 min of reaction while composite sample conversion starts under 90 % and decreases during the time. However the deactivation phenomenon is hindered by using core-shell catalyst; hence the robustness can be named as a privilege employing composite zeolitic catalysts.

## Introduction

Zeolites are the most widely used solid catalysts in industry due to their microporous structures and the resulting unique shape selectivity.Silicoalumophosphates (SAPO) are of high interest as acidic catalysts. Especially SAPO-34 gained a lot of attention due to the highpotential in the conversion of methanol to olefins[1-4]. However, the diameter of the narrow micropores in SAPO-34 are about 0.38 nm which restrict the intracrystalline diffusion of bulky molecules like carbocations as intermediate prodcuts. As a result, severe transport limitations negatively influence not only the reactivity but also both the selectivity and the lifetime. Therefore, the development of a facile route towards the formation of the hierarchical SAPO-34 and the correlation of its nanostructure and reactivity are important research topics[5]. In the petroleum industries, coke and asphaltenes aggregation[6, 7] contribute to catalyst deactivation[8, 9]. Furthermore, over the past decade, much effort has been devoted to prepare zeolites with enhanced accessibility of the reactant molecules to the active sites in order to achieve higher productyield/selectivity and to meet the undergoing commercial oil composition changes[10, 11]. Hierarchical zeolite is designated aszeolites with a bimodal distribution of porosity, which consequently exhibit reduced steric and diffusional restrictions[12].Recent progress has demonstrated that fabrication of core-shell composite materials have paid particular attention to the fields of catalysis due to their acidic properties of the external surface of zeolites and diffusion efficiency can be altered in a desired way[13]. In this contribution, our aim is to compare catalytic activity and life time of two novel catalysts in MTO process.Recently, two kind of mesoporous catalysts with high propylene selectivity in propane dehydrogenation (PDH) process have been successfully synthesized. The first catalyst is SAPO-34 molecular sieve with hierarchical tuned nanostructure[14, 15] which has been synthesized through a facile one-step hydrothermal route using a combination of amine agents [i.e., tetraethyl ammonium hydroxide (TEAOH), diethyl amine (DEA)] and polyethylene glycol (PEG), as structure directing and mesogenerating agents, respectively. The second catalyst is a novel bi-phase hierarchical SAPO-34/ZSM-5 zeolite composite[16] which consist of SAPO-34 TPA-exchanged robust core crystals and ZSM-5 shell around them.

## Experimental

### Synthesis of Catalysts

We prepared SAPO-34 molecular sieve and SAPO-34/ZSM-5 compositee zeolite according to our previous report [14, 16].

### *Catalytic test*

The MTO reactions were carried out in a packed-bed stainless steel reactor using 0.5 g ofcatalyst under the following conditions: methanol (MeOH) concentration of 30 wt.% in $H_2O$, 10 ml.min$^{-1}$ of $N_2$, reaction temperature of 673 K and WHSV of 3 h$^{-1}$. The reaction products were analyzed with an online GC (gas chromatography YL6100) equipped with HID detector.

## Results and discussion

### Characterization

XRD spectrograms of prepared molecular sieves along with reference components i.e. ZSM-5 and SAPO-34 are represented in Fig. 1 to detect the crystalline structure of the samples. As it can be observed, all typical characteristic peaks of both MFI (structural code for ZSM-5) and CHA (structural code for SAPO-34) structures are present in binary compositee sample and CHA structures are present in hierarchical SAPO-34. Also no other unidentified phase in XRD patterns is observed. In Overall, the diffraction peaks of binary composites are weaker compared to the pristine ZSM-5 and SAPO-34 based on the shielding effect of materials surrounding core seeds.

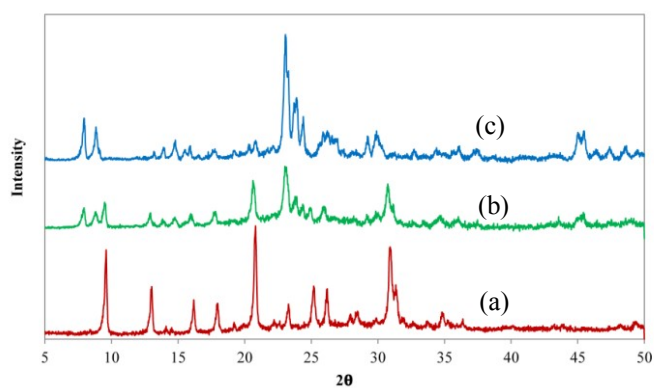

Fig. 1. XRD pattern of (a) hierarchical SAPO-34 [15], (b) SAPO-34/ZSM-5[16] (c) ZSM-5

FESEM image of hierarchical SAPO-34 shows unifrom cubic single crystals of about 3-4 μm.The composite photograph suggests huge change in themorphology of SAPO-34 cubic crystals.Semi-sphericalbig particles with rough surfaces were formed with ZSM-5 particles being attached and stacked on the surfaces of pre-treated SAPO-34 crystals.

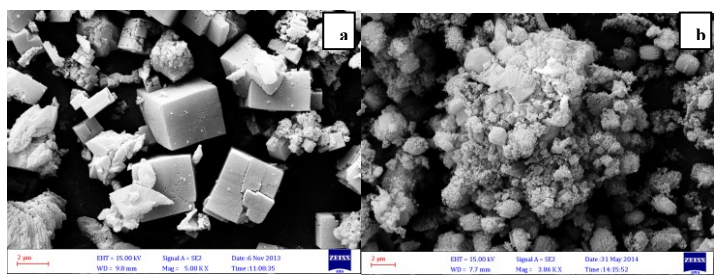

Fig. 2. FESEM images of a) hierarchical SAPO-34 and b) SAPO-34/ZSM-5

**Table 1**

Physicochemical properties of SAPO34, composites, and ZSM-5 zeolites.

| Sample | $S_{BET}$ (m$^2$/g) [a] | $S_{mic}$ (m$^2$/g) [b] | $V_t$ (cm$^3$/g) [c] | $V_{mic}$ (cm$^3$/g) [b] | Mean Pore Width (nm) [d] |
|---|---|---|---|---|---|
| SAPO-34 | 427.4 | 390 | 0.29 | 0.18 | 9.3 |
| SZ2 | 308.7 | 256.8 | 0.22 | 0.12 | 10.3 |
| ZSM-5 | 275.7 | 214.2 | 0.18 | 0.1 | 8.4 |

[a] $S_{BET}$ was obtained by analyzing nitrogen adsorption data at 77 K in a relative vapor pressure ranging from 0.05 to 0.3.
[b] Micropore area and micropore volume were determined using t-plot method.
[c] Total pore volume was estimated based on the volume adsorbed at P/P0~0.99.
[d] Mean pore width was determined by using BJH method.

Basic physicochemical and textural properties of obtained products are summarized in Table 1. BET surface area and pore volume of composite zeolitic molecular sieve is within the represented surface area and pore volume ranges of parent ZSM-5 and SAPO-34. Catalytic measurements. Mean pore width value of SZ2 sample is close to the relevant SAPO-34 material.

**Catalytic measurements**

The conversion and selectivity plots of the catalysts versus time on stream are shown in Fig 3 ,4, respectively, to characterize the catalytic performance of each catalyst.

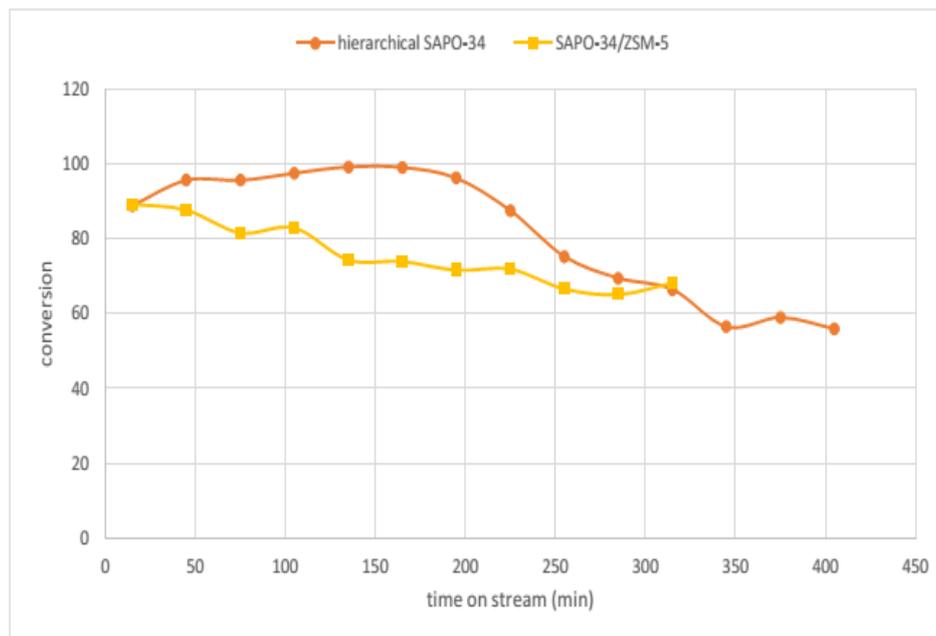

Fig. 3. Methanol conversion variation with time-on-stream over hierarchical SAPO-34 and SAPO-34/ZSM-5 catalysts

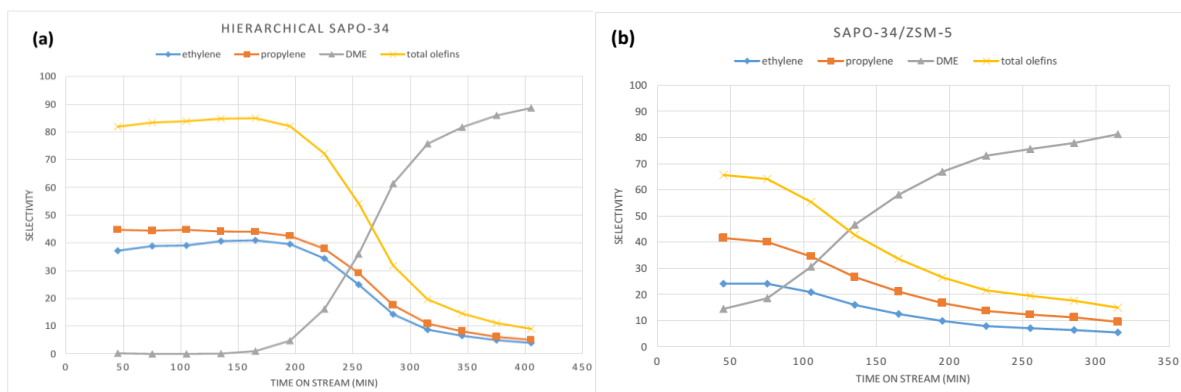

Fig. 4. The selectivity on (a) hierarchical SAPO-34 and (b) SAPO-34/ZSM-5 catalysts

As it can be seen, in Fig. 3 hierarchical SAPO-34 has a significant activity compared to SAPO-34/ZSM-5 composite. It shows a full conversion of MeOH during the first 200 min of reaction while composite sample conversion starts under 90 % and decreases during the time.Also, Fig. 4 shows that hierarchical SAPO-34 has a better olefins selectivity in comparison with the SAPO-34/ZSM-5. Most of the surface acid sites of composite are weak and moderate according to TPD profile[16] which favors the formation of alkanes and aromatics instead of alkenes. Simantenaously waek nature of acid sites promote the product diffusion and alleviate the coke deposition. As it can be observed, the deactivation rate of our composite catalyst is lower than SAPO-34, the conversion is not complete which is belived to be closely related to the non–protonic form of ZSM-5. We strongly believe that a simple ion-exchange process will create lots of strong Bronsted acidic centers which are essential to convert produced DME to light olefins to increase conversion and modify the product distribution.

**Conclusions**

The novel micro-meso porous zeolitic catalysts were synthesized for MTO process. The first catalyst was a kind of SAPO-34 molecular sieve with hierarchical tuned nanostructure, consisting micro and meso pores. The second one was a kind of bi-phase hierarchical SAPO-34/ZSM-5 zeolite composite consisting SAPO-34 TPA-exchanged robust core crystals with ZSM-5 shell. In this contribution, we compared their catalytic activity and life time in MTO process. Results show that hierarchical SAPO-34 has a significant lifetime and olefin selectivity compared with SAPO-34/ZSM-5. ZSM-5 has a relatively lower selectivity towards olefins, so when it is introduced as shell wrapping hierarchical SAPO-34 cores, its negative impact of selectivity can be predicted. However this negative impact can be evaded using an ion-exchange step to produce protonic form of ZSM-5 within the binary sructure.